\documentstyle[12pt,epsf]{article}
\newcommand{\bm}[1]{\mbox{\boldmath $#1$}}
\newcommand{\bt}[1]{{\bm #1}_{_T}}
\newcommand{\Pslash}{\kern 0.2 em P\kern -0.56em \raisebox{0.3ex}{/}}
\newcommand{\Sslash}{\kern 0.2 em S\kern -0.56em \raisebox{0.3ex}{/}}
\newcommand{\pslash}{\kern 0.2 em p\kern -0.4em /}
\newcommand{\kslash}{\kern 0.2 em k\kern -0.45em /}
\newcommand{\xbj}{x_{_B}}
\newcommand{\bkt}{\bt{k}}
\newcommand{\bpt}{\bt{p}}
\newcommand{\bqt}{\bt{q}}
\newcommand{\bSt}{\bt{S}}
\newcommand{\ba}{\begin{eqnarray}}
\newcommand{\ea}{\end{eqnarray}}
\newcommand{\be}{\begin{equation}}
\newcommand{\ee}{\end{equation}}
\begin{document}
\title{
\begin{flushright}
\small
NIKHEF 96-028\\
VUTH 96-07\\
nucl-th/9611040
\end{flushright}
Prospects for spin physics\\ in semi-inclusive processes\footnote{
Talk presented at the second ELFE Workshop, St. Malo, 23-27 September 1996.}
}
\author{
P.J. Mulders
\\
{\small
Department of Physics and Astronomy, Free University of Amsterdam}\\
{\small
and National Institute for Nuclear Physics and High-Energy Physics}\\
{\small
P.O. Box 41882, NL-1009 DB Amsterdam, the Netherlands}
}
\date{}
\maketitle
\begin{abstract}
I discuss inclusive and semi-inclusive lepton-hadron scattering
emphasizing the importance of polarization in order to study
various single or double spin asymmetries and the importance of
particle identification and angular resolution in the dectection of
final state particles to study azimuthal asymmetries. The observables
obtained in this way enable a detailed study of quark and gluon
correlations in hadrons.
%
%\begin{keyword}
%Polarized semi-inclusive deep inelastic scattering
%\end{keyword}
%
%\vspace{0.5 cm}
%PACS number(s): 
%
\end{abstract}
%
%\end{frontmatter}
%
%%%%%%%%%%%%%%%%%%%%%%%%%%%%%%%%%%%%%%%%%%%%%%%%%%%%%%
%
\section{Introduction}
The reason that spin has become a hot topic in deep inelastic scattering 
(DIS) is not only due to the progress in beam and target polarization 
enabling new experiments, but it is as important that there exists a 
clear view of what one actually measures. In particular this is the case
for sum rules 
that measure a well-determined local matrix element of quark and gluon 
field operators. For instance the integral $\int dx\ g_1^N(x)$ measures 
an axial current matrix element. For $\int dx\ (g_1^p-g_1^n)(x)$ this is 
even an combination that gives up to a factor the nucleon axial charge, a 
number that can also be obtained from neutron $\beta$-decay, a relation
known as the Bjorken sumrule.

The topic of this talk will be the relation of DIS observables to matrix 
elements of quark and gluon operators. In order to establish such a 
relation one needs to consider {\em hard} scattering processes in which a
large scale $Q$ -- in lepton-hadron scattering the four momentum transfer 
squared $-q^2 \equiv Q^2$ -- is present. The leading part of the cross 
section is related to matrix elements that have a simple interpretation 
as quark and gluon distributions. Subleading parts in an expansion in 
$1/Q$ involve matrix elements of combinations of quarks and gluon fields, 
i.e. quark-gluon correlation functions. These are known as {\em higher 
twist} contributions. A particular feature that I want to emphasize is 
the role of transverse momenta of quarks. To probe them one needs for 
example semi-inclusive processes such as 1-particle inclusive 
leptoproduction. The momentum of the produced hadron defines a direction 
orthogonal to the {\em fast} direction defined by the (spacelike) virtual 
photon and the target hadron. A similar role can be played by the spin 
vector if the target hadron is transversely polarized. Anyway as will 
become evident polarization plays an important role in this talk.

\begin{figure}[h]
\begin{center}
\leavevmode
\epsfxsize=5cm \epsfbox{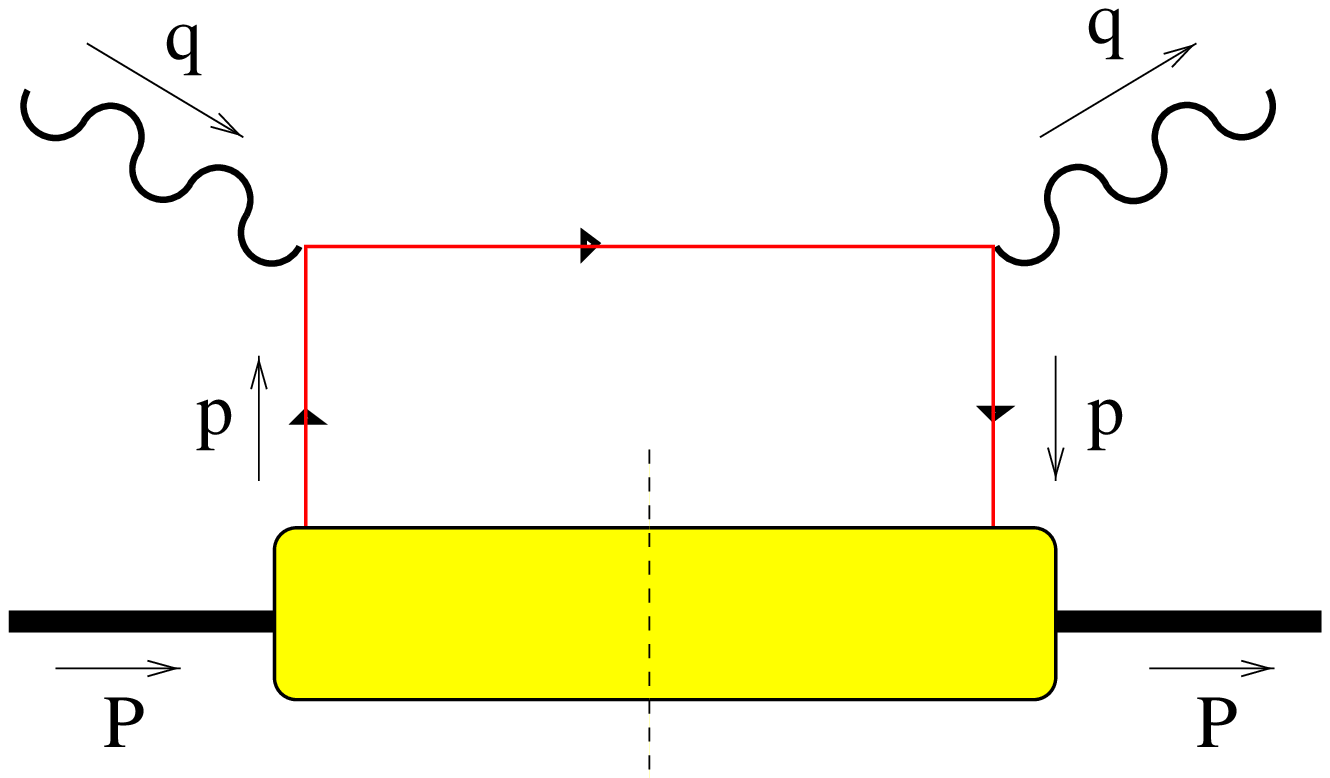}
\hspace{2 cm}
\epsfxsize=5cm \epsfbox{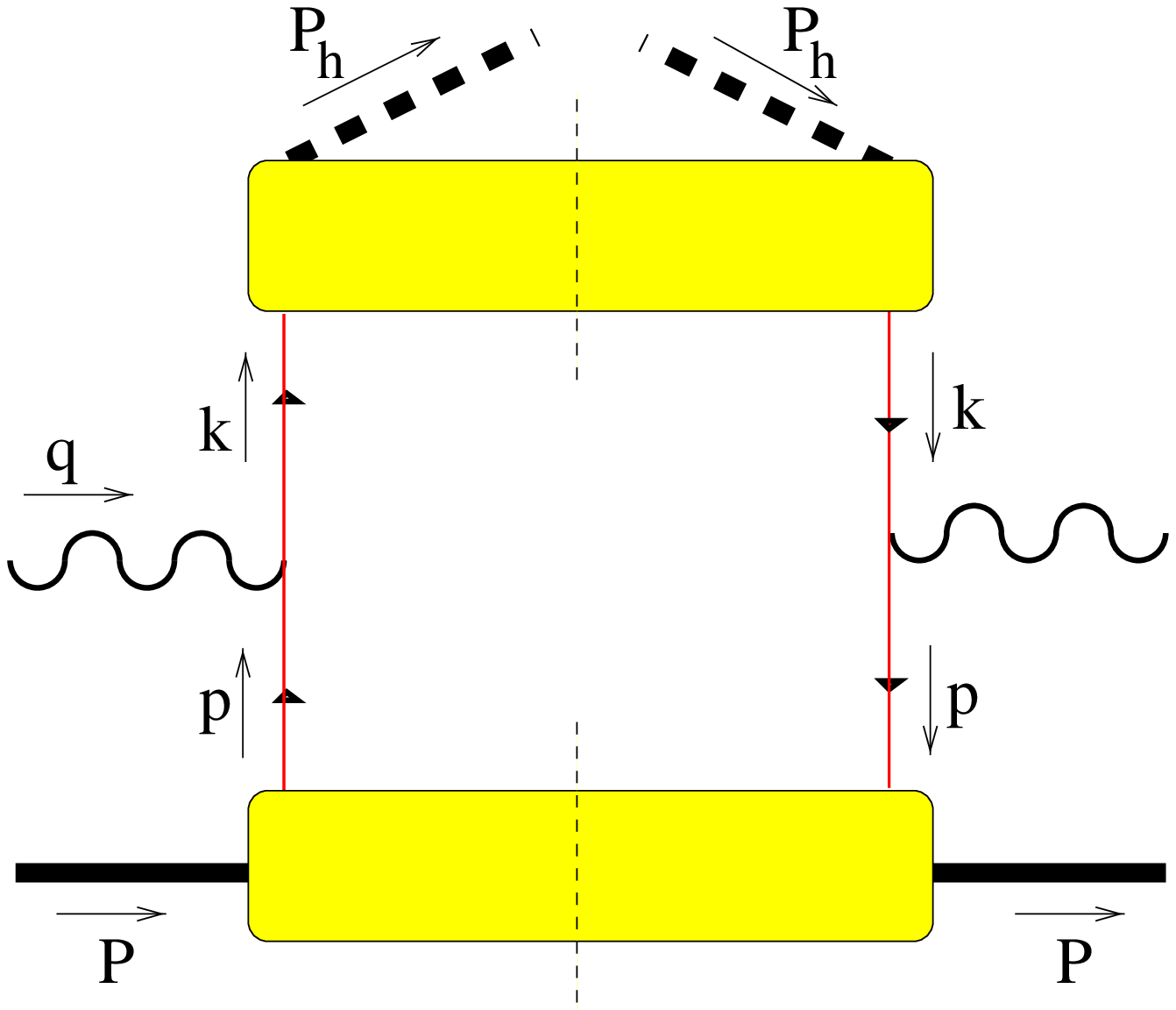}
\end{center}
\caption{\label{fig1}\em
The simplest (parton-level) diagrams representing the squared amplitude
in lepton hadron inclusive scattering (left) en semi-inclusive scattering
(right).}
\end{figure}
One of the aims of deep inelastic leptoproduction is the study of the 
quark and gluon structure of the hadronic target using the known 
framework of Quantum chromodynamics (QCD). Thus, as a theorist the aim is 
to calculate the hadronic tensor $W_{\mu\nu}$ by making a diagrammatic 
expansion. Already at the simplest level (Fig. 1) a problem is encountered, 
namely there are hadrons involved for which QCD does not provide rules. Thus, 
{\em soft parts} are identified that allow inclusion of hadrons in the 
field theoretical framework. Furthermore it will turn out that for $Q^2 
\rightarrow \infty$ only a limited number of diagrams is needed.

\section{Soft parts}

The soft parts can be written down in terms of quark and gluon fields
as illustrated below. They are characterized by the fact that 
the momenta are {\em soft} with respect to each other.
We have for the distribution part~\cite{Soper77,Jaffe83}

\begin{minipage}{6.5 cm}
\leavevmode
\epsfxsize=6 cm \epsfbox{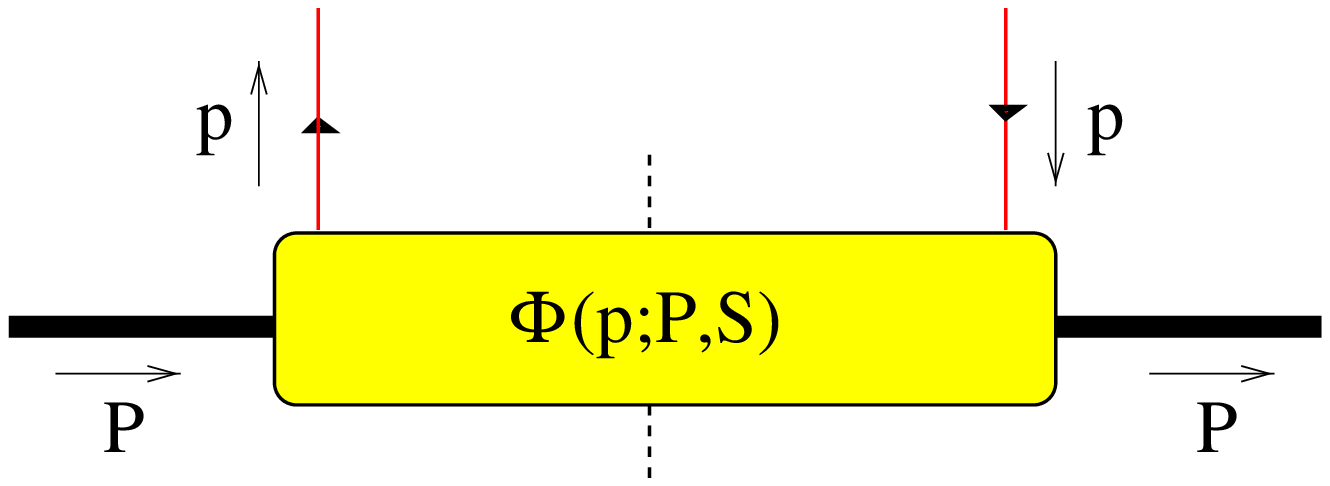}
\end{minipage}
\begin{minipage}{7.5 cm}
\[
\mbox{with}\ \ p^2 \sim p\cdot P \sim P^2 = M^2 \ll Q^2
\]
\end{minipage}
\newline
represented by
\ba
&&
\Phi_{ij}(p,P,S)=\frac{1}{(2\pi)^4}\int d^4x \;
e^{ip\cdot x}\;\langle P,S|\overline\psi_j(0)\psi_i(x)|P,S\rangle,
\ea
and the fragmentation part~\cite{CS82} 

\begin{minipage}{6.0 cm}
\leavevmode
\epsfxsize=4.5 cm \epsfbox{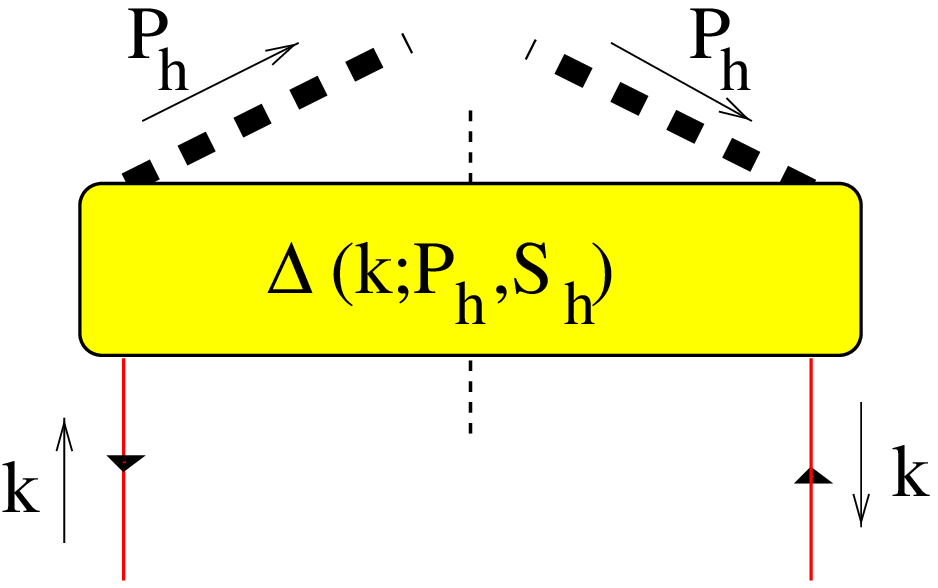}
\end{minipage}
\begin{minipage}{8 cm}
\[
\mbox{with}\ \ k^2 \sim k\cdot P_h \sim P_h^2 = M_h^2 \ll Q^2
\]
\end{minipage}
\newline
represented by
\ba
&&\Delta_{ij}(k,P_h,S_h)
\nonumber \\&&\qquad =\sum_X\frac{1}{(2\pi)^4}\int d^4x\;
e^{ik\cdot x} \langle 0|\psi_i(x)|P_h,S_h;X\rangle
\langle P_h,S_h;X|\overline\psi_j(0)|0\rangle.
\ea

In order to find out which information in the soft parts $\Phi$ 
and $\Delta$ is important in a hard process one needs to realize 
that the hard scale $Q$ leads in a natural way to the use of lightlike
vectors $n_+$ and $n_-$ satisfying $n_+^2 = n_-^2 = 0$ and $n_+\cdot n_-$
= 1. For inclusive deep inelastic scattering\footnote{
the lightlike vectors may be rescaled into dimensionful
vectors $\hat p$ = $(Q/\xbj\sqrt{2})n_+$ and $\hat n$ = $(\xbj\sqrt{2}/Q)n_-$,
thus
%\begin{eqnarray*}
%&&
\[
P = \hat p + \frac{M^2}{2}\,\hat n
%\\
\qquad \mbox{and} \qquad
%&&
q = -\xbj \,\hat p + \frac{Q^2}{2\xbj}\,\hat n .
%\hspace{3 cm} \mbox{}
%\end{eqnarray*}
\]
}
\[
\left.
\begin{array}{l} q^2 = -Q^2 \\
P^2 = M^2 \\
2\,P\cdot q = \frac{Q^2}{\xbj} \end{array} \right\}
\longleftrightarrow \left\{
\begin{array}{l}
P = \frac{\xbj M^2}{Q\sqrt{2}}\,n_-
+ \frac{Q}{\xbj \sqrt{2}}\,n_+
\\ \mbox{} \\
q = \frac{Q}{\sqrt{2}}\,n_-
- \frac{Q}{\sqrt{2}}\,n_+
\end{array} \right.
\]

Comparing the power of $Q$ with which the momenta in the soft and hard
part appear one immediately is led to the fact that 
$\int dp^-\,\Phi(p,P,S)$ is the relevant quantity to investigate,

\begin{minipage}{5.5 cm}
%\leavevmode
\epsfxsize=5cm \epsfbox{mulders1.eps}
\end{minipage}
\hspace{1.5 cm}
\begin{minipage}{7 cm}
\begin{tabular}{c|cc|c}
part & \multicolumn{2}{c}{'components'} & \\ 
& $-$ & + &  \\
\hline
HARD &  $\sim Q$  &   $\sim Q$  & \\
$H\rightarrow q$ & $\sim 1/Q$  &  $\sim Q$  &
$\rightarrow \int dp^- \ldots$
\end{tabular}
\end{minipage}

For 1-particle inclusive scattering one parametrizes the momenta as
\[
\left.
\begin{array}{l} q^2 = -Q^2 \\
P^2 = M^2\\
P_h^2 = M_h^2 \\
2\,P\cdot q = \frac{Q^2}{\xbj} \\
2\,P_h\cdot q = -z_h\,Q^2
%P\cdot P_h = z_h\,P\cdot q
\end{array} \right\}
\longleftrightarrow \left\{
\begin{array}{l}
P_h = \frac{z_h\,Q}{\sqrt{2}}\,n_-
+ \frac{M_h^2}{z_h\,Q\sqrt{2}}\,n_+
\\ \mbox{} \\
q =\frac{Q}{\sqrt{2}}\,n_- - \frac{Q}{\sqrt{2}}\,n_+ + q_T
\\ \mbox{} \\
P = \frac{\xbj M^2}{Q\sqrt{2}}\,n_-
+ \frac{Q}{\xbj \sqrt{2}}\,n_+
\end{array}
\right.
\]
and it follows that one needs besides the quantity $\int dp^-\,\Phi(p,P,S)$ 
the equivalent for the fragmentation part, $\int dk^+\,\Delta(k,P_h,S_h)$.

\begin{minipage}{5.5 cm}
%\leavevmode
\epsfxsize=5cm \epsfbox{mulders2.eps}
\end{minipage}
\hspace{1.5 cm}
\begin{minipage}{7 cm}
\begin{tabular}{c|cc|c}
part & \multicolumn{2}{c}{'components'} & \\ 
& $-$ &  + &  \\
\hline
$q\rightarrow h$ & $\sim Q$  &  $\sim 1/Q$  &
$\rightarrow \int dk^+ \ldots$ \\
HARD &  $\sim Q$  &   $\sim Q$  & \\
$H\rightarrow q$ & $\sim 1/Q$  &  $\sim Q$  &
$\rightarrow \int dp^- \ldots$
\end{tabular}
\end{minipage}

\section{Analysis of soft parts: distribution and fragmentation functions}

Hermiticity, parity and time reversal invariance (T) constrain the quantity
$\Phi(p,P,S)$. It must be of the form~\cite{RS79,TM95a}
\ba
\Phi(k,P,S) & = &
M\,A_1 + A_2\,\Pslash + A_3\kslash
+ i\,A_4\,\frac{[\Pslash,\kslash]}{2M}
+ i\,A_5\,(k\cdot S) \gamma_5
\nonumber \\ & &
+ M\,A_6 \,\Sslash \gamma_5
+ A_7\,\frac{k\cdot S}{M}\,\Pslash \gamma_5
+ A_8\,\frac{k\cdot S}{M} \kslash \gamma_5
+ A_9\,\frac{[\Pslash,\Sslash]\gamma_5}{2}
\nonumber\\ & &
+ A_{10}\,\frac{[\kslash,\Sslash]\gamma_5}{2}
+ A_{11}\,\frac{k\cdot S}{M}\,\frac{[\Pslash,\kslash]\gamma_5}{2M}
+ A_{12}\, \frac{\epsilon_{\mu \nu \rho \sigma}\gamma^\mu P^\nu
k^\rho S^\sigma}{M},
\ea
with real $A_i$ = $A_i(k^2,k\cdot P)$ and if T applies $A_4 = A_{11} = A_{12}
= 0$.

This imposes constraints on the functions allowed in the Dirac projections
$\Phi^{[\Gamma]}$ defined as
\begin{eqnarray}
\Phi^{[\Gamma]}(x,\bm p_T) & = &
\int dp^- \,\frac{Tr[\Phi \Gamma]}{2}
\nonumber \\ & = &
\left. \int \frac{d\xi^-d^2\bm \xi_T}{2\,(2\pi)^3}\ e^{ip\cdot \xi}
\,\langle P,S\vert \overline \psi(0) \Gamma \psi(\xi)
\vert P,S\rangle \right|_{\xi^+ = 0},
\end{eqnarray}
which is a lightfront ($\xi^+$ = 0) correlation function. 
The relevant projections in $\Phi$ that are important in leading order in 
$1/Q$ in hard processes are
\begin{eqnarray}
\Phi^{[\gamma^+]}(x,\bpt) & = &
f_1(x ,\bpt^2)
\\ 
\Phi^{[\gamma^+ \gamma_5]}(x,\bpt) & = &
\lambda\,g_{1L}(x ,\bpt^2)
+ \frac{(\bpt\cdot\bSt)}{M}\,g_{1T}(x ,\bpt^2)
\\ 
\Phi^{[ i \sigma^{i+} \gamma_5]}(x,\bpt) & = &
S_T^i\,h_1(x ,\bpt^2)
+ \frac{\lambda\,p_T^i}{M} \,h_{1L}^\perp(x ,\bpt^2)
\nonumber \\ && {}
- \frac{\left(p_T^i p_T^j + \frac{1}{2}\bpt^2g_T^{ij}\right)
S_{Tj}}{M^2}\,h_{1T}^\perp(x ,\bpt^2)
\end{eqnarray}
Here $x = p^+/P^+$ and the (lightcone) helicity $\lambda$ and transverse
spin $S_T$ of the target hadron are defined as
\begin{eqnarray*}
\left.
\begin{array}{l} 
P^2 = M^2\\
S^2 = -1 \\
P\cdot S = 0 
\end{array} \right\}
&\longleftrightarrow & \left\{
\begin{array}{l}   
P = \frac{M^2}{2P^+}\,n_-
+ P^+\,n_+ \\
S = -\lambda\frac{M}{2P^+}\,n_-
+ \lambda\frac{P^+}{M}\,n_+ + S_T
\end{array}
\right.   
\\&& \quad (\lambda^2 + \bm S_T^2 = 1)
\end{eqnarray*}

\begin{figure}[b]
\leavevmode
\epsfxsize=2.1 cm \epsfbox{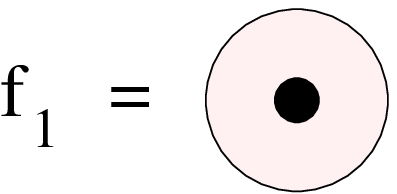}
\newline
\epsfxsize=5.0 cm \epsfbox{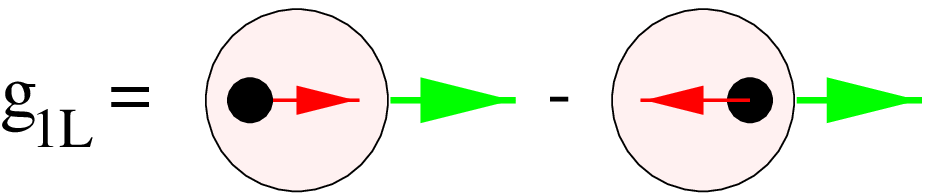}
\hspace{0.5 cm}
\epsfxsize=3.7 cm \epsfbox{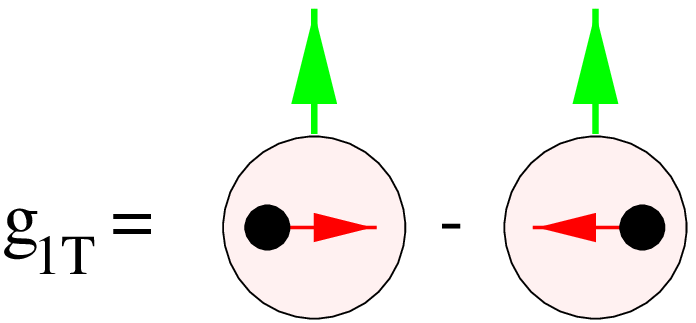}
\newline
\epsfxsize=3.7 cm \epsfbox{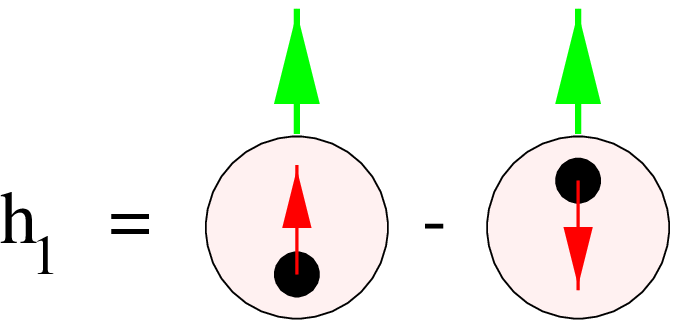}
\hspace{0.5 cm}
\epsfxsize=5.0 cm \epsfbox{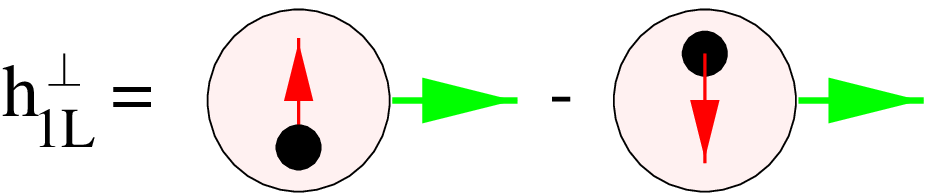}
\hspace{0.5 cm}
\epsfxsize=3.7 cm \epsfbox{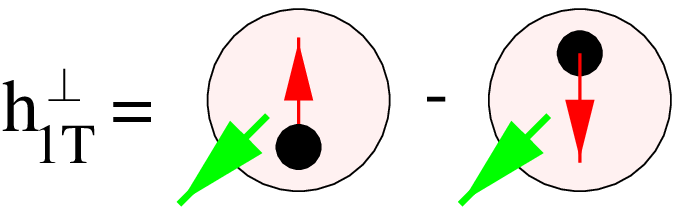}
\caption{\label{fig2}\em
The interpretation of the leading twist functions appearing in the
Dirac projections of the soft part $\Phi$.}
\end{figure}
All functions appearing above can be interpreted as momentum space densities,
as illustrated in Fig. 2.
The first, denoted $f_{\ldots}^{\ldots}$ 
(in naming we follow ref.~\cite{JJ92}) involve the operator structure
\ba
&&
\Phi^{[\gamma^+]} \propto \overline \psi \gamma^+ \psi = \psi_+^\dagger
\psi_+,
\ea
where $\psi_+ = P_+\psi$ with $P_+ = \gamma^-\gamma^+/2$. This operator
projects on the socalled good component of the Dirac field, which can be
considered as a {\em free} dynamical degree of freedom in front form
quantization. It is precisely in this sense that partons measured in
hard processes are free. The functions $g_{\ldots}^{\ldots}$ and
$h_{\ldots}^{\ldots}$ appearing above are differences of densities
for opposite quark polarizations. In the case of $g_{\ldots}^{\ldots}$ 
it is the difference of densities for quark chirality states
$\psi_{R/L} = P_{R/L}\psi$ with $P_{R/L} = (1\pm \gamma_5)/2$,
\ba
&&
\Phi^{[\gamma^+\gamma_5]} \propto \overline 
\psi \gamma^+\gamma_5 \psi = 
\psi_{+R}^\dagger \psi_{+R} - \psi_{+L}^\dagger \psi_{+L},
\ea
while in the case of $h_{\ldots}^{\ldots}$ it is the
difference of densities for quark transverse spin states
$\psi_{\uparrow/\downarrow} = P_{\uparrow/\downarrow}\psi$ 
with $P_{\uparrow/\downarrow} = (1\pm \gamma^1\gamma_5)/2$,
\ba
&&
\Phi^{[\sigma^{1+}\gamma_5]} \propto \overline 
\psi \sigma^{1+}\gamma_5 \psi = 
\psi_{+\uparrow}^\dagger \psi_{+\uparrow} 
- \psi_{+\downarrow}^\dagger \psi_{+\downarrow}.
\ea
The projectors $P_{R/L}$ and $P_{\uparrow/\downarrow}$ commute with $P_+$.

Theorists often are more happy with lightcone correlation functions,
because they can employ and isolate lightcone singularities and proof
factorization theorems. These appear in the $\bpt$-integrated quantities
\ba
\Phi^{[\Gamma]}(x) & = &
\int dp^-\,d^2\bm p_T \,\frac{Tr[\Phi \Gamma]}{2}
\nonumber \\ & = &
\left. \int \frac{d\xi^-}{4\pi}\ e^{ip\cdot \xi}
\,\langle P,S\vert \overline \psi(0) \Gamma \psi(\xi)
\vert P,S\rangle \right|_{\begin{array}{l} \xi^+ = \vert\bm \xi_T\vert = 0 
\end{array}}
\ea
The above constraints on $\xi$ imply $\xi^2 = 0$. The $\bpt$-averaged
results which are relevant for inclusive lepton-hadron or $\bqt$-averaged
Drell-Yan are
\ba
& & \Phi^{[\gamma^+]}(x) =
f_1(x),
\\ & & \Phi^{[\gamma^+ \gamma_5]}(x) =
\lambda\,g_{1}(x),
\\ & & \Phi^{[ i \sigma^{i+} \gamma_5]}(x) =
S_T^i\,h_{1}(x).
\ea
It is useful to remark here that flavor indices have been omitted, i.e.
one has $f_1^u$, $f_1^d$, etc. At this point it may also be good to mention
other notations used frequently such as $f_1^u(x) = u(x)$, $g_1^u(x) =
\Delta u(x)$, $h_1^u(x) = \Delta_T u(x)$, etc. We note that some of the
$\bpt$-dependent functions have vanished. They do appear, however, in
$\bpt/M$-weighted results relevant for azimuthal asymmetries in 1-particle
inclusive $\ell H$ or DY,
\ba
\Phi_\partial^{\alpha[\Gamma]}(x) & \equiv &
\int dp^-\,d^2\bm p_T \ p_T^\alpha\frac{Tr[\Phi \Gamma]}{2}
\nonumber \\ & = &
\left. \int \frac{d\xi^-}{4\pi}\ e^{ip\cdot \xi}
\,\langle P,S\vert \overline \psi(0) \Gamma \,i\partial^\alpha\psi(\xi)
\vert P,ngle \right|_{\begin{array}{l}
\xi^+ = \vert\bm \xi_T\vert = 0 \end{array} }
\ea
Two $\bpt/M$-weighted results (relevant for asymmetries in $\ell H$ or DY)
are
\ba
&&\frac{1}{M}\,\Phi_\partial^{\alpha[\gamma^+ \gamma_5]}(x) =
S_T^\alpha \,g_{1T}^{(1)}(x),
\\
&&\frac{1}{M}\,\Phi_\partial^{\alpha[ i\sigma^{i+}\gamma_5]}(x) =
-g_T^{\alpha i}\,\lambda\,h_{1L}^{\perp (1)}(x),
\ea
where the functions with superscript $(1)$ denote $\bpt^2$-moments,
\ba
&&
g_{1T}^{(1)}(x) \equiv
\int d^2\bpt\,\frac{\bpt^2}{2M^2}\,g_{1T}(x ,\bpt).
\ea

The analysis of the soft part $\Phi$ can be extended to other Dirac
projections. Limiting ourselves to $\bpt$-averaged functions one finds
the following possibilities,
\ba
& & \Phi^{[1]}(x) =
\frac{M}{P^+}\,e(x),
\\ & & \Phi^{[ \gamma^i \gamma_5]}(x) =
\frac{M\,S_T^i}{P^+} \, g_{T}(x),
\\ & & \Phi^{[ i\sigma^{+-}\gamma_5 ]}(x) =
\frac{M}{P^+}\,\lambda\,h_{L}(x).
\ea
Lorentz covariance requires for these projections on the right hand side
a factor $M/P^+$, which as can be seen from the earlier given parametrization
of momenta produces a suppression factor $M/Q$ and thus these functions
appear at subleading order in cross sections. Furthermore, these functions
can not be written in the form of densities or difference of densities.
However, interesting relations between these functions and the above
$\bpt/M$-weighted functions can be obtained~\cite{BKL84,TM96} using the most 
general
amplitude analysis for $\Phi$, constrained by hermiticity, parity and
time reversal invariance,
\ba
&&
g_2 \ =\  g_T - g_1 \ =\  \frac{d}{dx}\,g_{1T}^{(1)},
\\ &&
h_2\ =\ 2(h_L - h_1) \ =\  -2\,\frac{d}{dx}\,h_{1L}^{\perp(1)}.
\ea

Just as for the distribution functions one can perform an analysis of
the soft part describing the quark fragmentation~\cite{TM96}. 
The Dirac projections in this case are
\ba
\Delta^{[\Gamma]}(z,\bkt) & = &
\int dk^+\,\frac{Tr[\Delta\Gamma]}{4z}
\nonumber \\ & = &
\left. \sum_X \int \frac{d\xi^+d^2\bm \xi_T}{4z\,(2\pi)^3} \,
e^{ik\cdot \xi} \,Tr  \langle 0 \vert \psi (x) \vert P_h,X\rangle
\langle P_h,X\vert\overline \psi(0)\Gamma \vert 0 \rangle
\right|_{\xi^- = 0}.
\ea
The relevant projections in $\Delta$ that appear in leading order in 
$1/Q$ in hard processes are for the case of no final state polarization,
\ba
& & \Delta^{[\gamma^-]}(z,\bkt) =
D_1(z,-z\bkt),
\\ & & \Delta^{[i \sigma^{i-} \gamma_5]}(z,\bkt) =
\frac{\epsilon_T^{ij} k_{T j}}{M_h}\,H_1^\perp(z,-z\bkt).
\qquad \mbox{[T-odd]}
\ea
The arguments of the fragmentation functions $D_1$ and $H_1^\perp$ are 
chosen to be $z$ = $P_h^-/k^-$ and $\bm P_{h\perp}$ = $-z\bkt$. The first 
is the (lightcone) momentum fraction of the produced hadron, the second 
is the transverse momentum of the produced hadron with respect to the quark.
The fragmentation function $D_1$ is the equivalent of the distribution 
function $f_1$. It can be interpreted as the probability of finding a 
hadron $h$ in a quark. Adding quark flavor index ($a$) and kind of 
produced hadron the distribution functions are normalized as
$\sum_h \int dz\,d^2P_{h\perp}\,zD_1(z,P_{h\perp})$ = 1. Noteworthy is the
appearance of the function $H_1^\perp$, interpretable as the different 
production probability of unpolarized hadrons from a transversely 
polarized quark (see Fig. 3). This functions has no equivalent in the 
distribution functions and is allowed because of the non-applicability of 
time reversal invariance because of the  appearance of out-states 
$\vert P_h, X\rangle$ in $\Delta$, rather than the plane 
wave states in $\Phi$.
\begin{figure}[t]
\leavevmode
\epsfxsize=2.1 cm \epsfbox{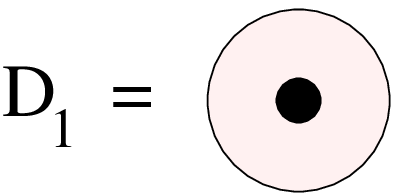}
\hspace{0.5 cm}
\epsfxsize=3.7 cm \epsfbox{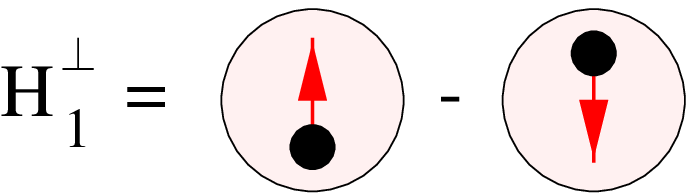}
\caption{\label{fig3}\em
The interpretation of the leading twist functions appearing in the
Dirac projections of the soft part $\Delta$ for unpolarized hadrons.}
\end{figure}

After $\bkt$-averaging one is left with the function
\ba
& & \Delta^{[\gamma^-]}(z) =
D_1(z),
\ea
while the $\bkt/M$-weighted result is
\ba
&&\frac{1}{M}\,\Delta_\partial^{\alpha[ i\sigma^{i-}\gamma_5]}(z) =
-\epsilon_T^{\alpha i}\,H_1^{\perp (1)}(z),
\ea
with
\ba
&&H_1^{\perp(1)}(z) \equiv
z^2\int d^2\bkt\,\frac{\bkt^2}{2M_h^2}\,H_1(z ,-z\bkt).
\ea
At subleading order in hard processes one finds after $\bkt$-averaging 
the following fragmentation functions
\ba
& & \Delta^{[1]}(z) =
\frac{M_h}{P_h^-}\,E(z),
\\ & & \Delta^{[ i\sigma^{ij}\gamma_5 ]}(z) =
\frac{M_h\,\epsilon_T^{ij}}{P_h^-}\,H(z).
\ea
Again a relation with the $\bkt/M$-weighted functions exist, namely
\ba
&&\frac{H(z)}{z}\ =\  z^2\,\frac{d}{dz}\,\left[\frac{H_1^{\perp (1)}}{z}
\right].
\ea
We summarize the analysis of the soft part with a table of distribution and
fragmentation functions for unpolarized (U), longitudinally polarized (L) 
and transversely polarized (T) targets, distinguishing leading (twist 
two) and subleading (twist three, appearing at order $1/Q$) functions and 
furthermore distinguishing the chirality. Chiral even functions are 
diagonal in the space of chiral quark states, which is the case for Dirac 
projections $\gamma^\mu$ or $\gamma^\mu\gamma_5$ (thus the functions 
$f$, $g$, $D$ and $G$); 
chiral odd functions are nondiagonal in this basis, which is the case for 
Dirac projections $1$, $i\gamma_5$ or $i\sigma^{\mu\nu}\gamma_5$ (thus 
the functions $e$, $h$, $E$ and $H$). 
The functions printed in boldface survive after integration over transverse
momenta.
\vspace{0.2cm}
\newline
\begin{minipage}{5.6 cm}
\begin{center}
\begin{tabular}{|c|r|c|c|} \hline
\multicolumn{4}{|c|}{DISTRIBUTIONS} \\ \hline
\multicolumn{2}{|c|}{} & \multicolumn{2}{c|}{chirality} \\
\multicolumn{2}{|c|}{$\Phi^{[\Gamma]}$} & even & odd \\ \hline
&  {U} & ${\bm f}_1$ & \\
twist 2 & {L} & ${\bm g}_{1L}$ & $h_{1L}^\perp$ \\
&  {T} & $g_{1T}$ & ${\bm h}_1\ \ {h_{1T}^\perp}$  \\ \hline
&  {U} & ${f^\perp}$ & ${\bm e}$ \\
twist 3 & {L} & ${g_L^\perp}$ & ${\bm h}_L$ \\
&  {T} & ${\bm g}_T\ \ {g_T^\perp}$ &
${h_T}\ \ {h_T^\perp}$  \\
\hline
\end{tabular}
\end{center}
(9 independent functions)
\end{minipage}
\begin{minipage}{8.4 cm}
\begin{center}
\begin{tabular}{|c|r|c|c|} \hline
\multicolumn{4}{|c|}{FRAGMENTATION} \\ \hline
\multicolumn{2}{|c|}{} & \multicolumn{2}{c|}{chirality} \\
\multicolumn{2}{|c|}{$\Delta^{[\Gamma]}$} & even & odd \\ \hline
&  {U} & ${\bm D}_1$ & ${H_{1}^\perp}$\\
twist 2 & {L} & ${\bm G}_{1L}$ & ${H_{1L}^\perp}$ \\
&  {T} & ${G_{1T}}\ \ {D_{1T}^\perp}$ & ${\bm H}_1\ \ {H_{1T}^\perp}$ \\
\hline &  {U} & ${D^\perp}$ & ${\bm E}\ \ {\bm H}$ \\
twist 3 & {L} & ${G_{L}^\perp}\ \ {D_L^\perp}\ \ {\bm E}_L$ & ${\bm H}_L$ \\
&  {T} & ${\bm G}_T\ \ {G_T^\perp}\ \ {\bm D}_T\ \ {E_T}$ &
${H_T}\ \ {H_T^\perp}$  \\
\hline \end{tabular}
\end{center}
(12 independent functions)
\end{minipage}
\newline
Note that the number of independent 
functions is determined by the number of amplitudes $A_i$ in the 
amplitude analysis, i.e. 9 and 12 for distribution and fragmentation 
functions respectively.

\section{Cross sections for lepton-hadron scattering}

Having completed the analysis of the soft parts, the next step is to find 
out how one obtains the information on the various correlation functions 
from experiments. We will limit ourselves here to lepton-hadron scattering
via one-photon exchange. The relevant kinematic (scaling) variables are 
given in the figure.
\newline
\begin{minipage}{10 cm}
%\leavevmode
\epsfxsize=8.5 cm \epsfbox{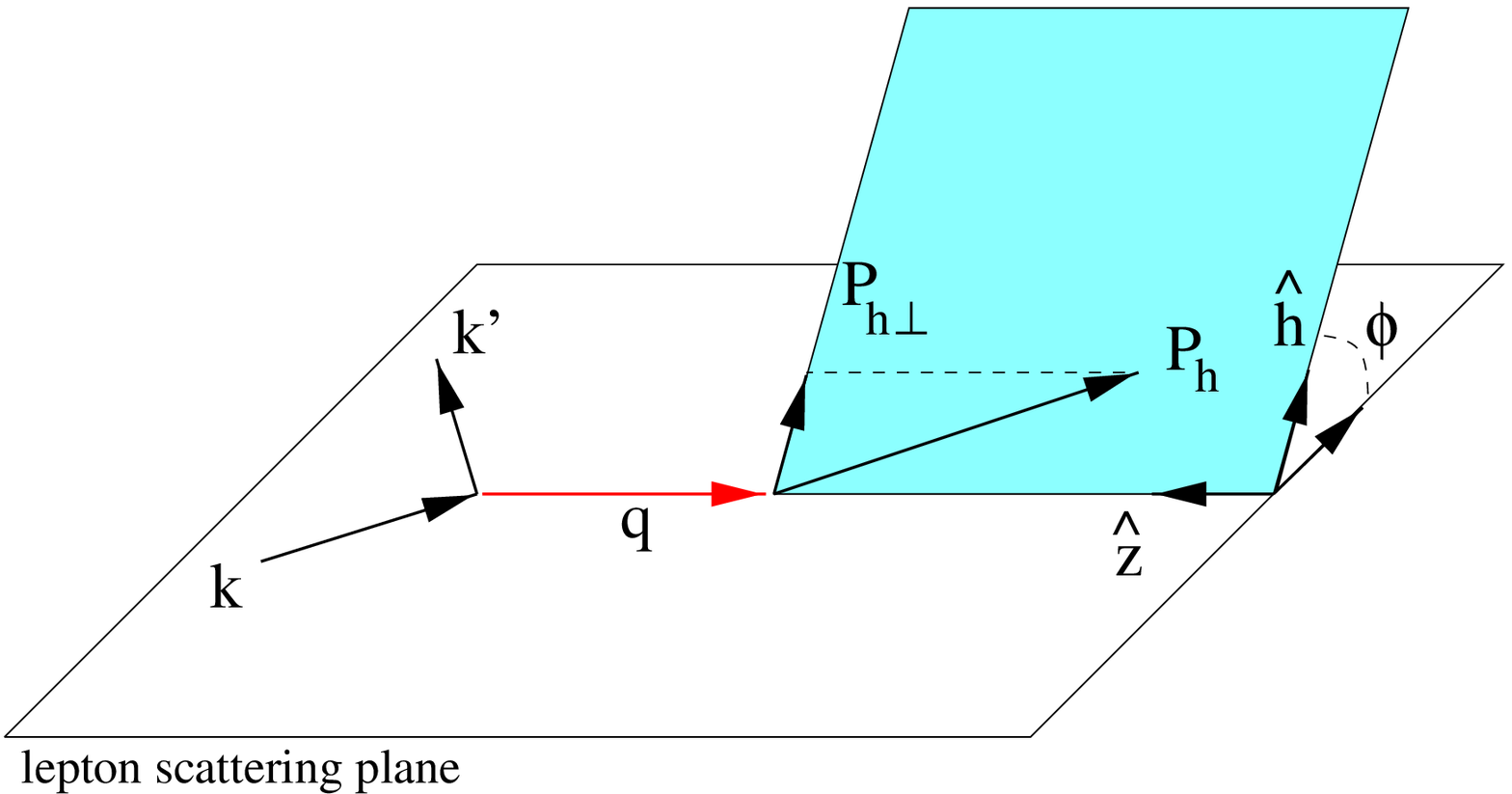}
\end{minipage}
\begin{minipage}{4 cm}
\begin{eqnarray*}
&&\ell H \longrightarrow \ell^\prime h X
\end{eqnarray*}
\begin{eqnarray*}
&&\xbj = \frac{Q^2}{2P\cdot q} \\
&&y = \frac{P\cdot q}{P\cdot k}\\
&&z_h = \frac{P\cdot P_h}{P\cdot q}
\end{eqnarray*}
\end{minipage}
\newline
To get the leading order result for semi-inclusive scattering it is 
sufficient to compute the diagram in Fig. 1 (plus the antiquark 
analogue) by using QCD and QED Feynman rules in the hard part and the 
matrix elements $\Phi$ and $\Delta$ for the soft parts, parametrized in 
terms of distribution and fragmentation functions. The results are:
\newline\newline
\fbox{
\begin{minipage}{13.0 cm}
{\bf Cross sections (leading in $1/Q$)}
\ba
&&\frac{d\sigma_{OO}}{d\xbj\,dy\,dz_h}
= \frac{2\pi \alpha^2\,s}{Q^4}\,\sum_{a,\bar a} e_a^2
\left\lgroup 1 + (1-y)^2\right\rgroup \xbj {f^a_1}(\xbj)\,{ D^a_1}(z_h)
\\ && \frac{d\sigma_{LL}}{d\xbj\,dy\,dz_h}
= \frac{2\pi \alpha^2\,s}{Q^4}\,{\lambda_e\,\lambda}
\,\sum_{a,\bar a} e_a^2\  y (2-y)\  \xbj {g^a_1}(\xbj)\,{D^a_1}(z_h)
\ea
\end{minipage}
}
\newline\newline
One recognizes in these well-known cross sections the characteristic 
structure of the 1-particle inclusive cross sections, namely the cross 
section $2\pi\,\alpha^2\,s/Q^4$ multiplying a kinematic 
factor (depending on $y$) and a sum over flavors of a product of 
a distribution function depending on $\xbj$ and a fragmentation function 
depending on $z_h$.

Some general rules on where to find a specific correlation function are
the following:
\begin{itemize}
\item
Depending on the twist $t$ of correlation functions, they appear in
cross sections suppressed by a factor $\propto (1/Q)^{t-2}$.
\item
Cross sections are chirally even. This means e.g. that one finds 
combinations like $f_1\,D_1$, ${h_1}\,{H_1}$ or ${e}\,{H_1^\perp}$. 
However, quark
mass terms are chirally odd, i.e. one may encounter combinations
like $m\,{f_1}\,{H_1^\perp}$.
\item
The number of spin vectors is even in case of a T-even fragmentation 
function, i.e. there are only double spin asymmetries, such as
${\lambda_e\,\lambda}\,{g_1}\,{D_1}$, ${\lambda_e\,\vert \bm S_T \vert}
\,{g_{1T}^{(1)}}\,{D_1}$ or ${\vert \bm S_T\vert \,\vert \bm S_{hT}\vert}
\,{h_1}\,{H_1}$.
\item
The number of spin vectors is odd in case of a T-odd fragmentation 
function, i.e. there are only single spin asymmetries, such as
${\lambda_e}\,{e}\,{H_1^\perp}$, ${\lambda}\,{h_{1L}^{\perp (1)}}\,{H_1^\perp}$ 
or ${\vert \bm S_{hT}\vert}\,{f_1}\,{D_T}$.
\end{itemize}

Two examples of subleading cross sections\footnote{
in all cases where we give polarized cross sections it should 
be realized that the full expression is $\sigma = \sigma_{OO} + \sigma_{OT} 
+ \ldots$, i.e. one needs to consider a difference of cross sections to 
isolate $\sigma_{OT}$
} 
are~\cite{TM96,JJ93}
\newline\newline
\fbox{
\begin{minipage}{13.0 cm}
{\bf Cross sections subleading in $1/Q$}
\ba
&&
\frac{2\pi\,d\sigma_{{OT}}}{d\xbj\,dy\,dz_h\,d\phi_e}
= \frac{2\pi \alpha^2\,s}{Q^4}\,{\vert \bm S_T\vert}
\,\sin(\phi_S) \,2(2-y)\sqrt{1-y}
\nonumber \\ &&
\hspace{4 cm} \mbox{}\times \sum_{a,\bar a} e_a^2
\,\frac{2M}{Q}\,\xbj {h^a_1}(\xbj)\,\frac{{\tilde H^a}(z_h)}{z_h},
\\ &&
\mbox{note:}
\ \frac{{\tilde H^a}(z)}{z}
= \frac{H^a(z)}{z} + 2\,H_1^{\perp (1)\,a}(z)
= \frac{d}{dz}\left[ z\,H_1^{\perp (1)\,a}\right],
\nonumber
\ea
\ba
&&
\frac{2\pi\,d\sigma_{{LT}}}{d\xbj\,dy\,dz_h\,d\phi_e}
= \frac{2\pi \alpha^2\,s}{Q^4}\,{\lambda_e \vert \bm S_T\vert}
\,\cos(\phi_S)\,2y\sqrt{1-y}
\nonumber \\ && 
\hspace{1 cm} \mbox{}\times \sum_{a,\bar a} e_a^2
\Biggl\{
\frac{2M}{Q}\,\xbj^2 {g_T^a}(\xbj)\,{D_1^a}(z_h)
%\\ && \quad\qquad \mbox{} 
+ \frac{2M_h}{Q}\,\xbj {h_1^a}(\xbj)\,\frac{{\tilde E^a}(z_h)}{z_h}
\Biggr\},
\\ &&
\mbox{note:}
\ \frac{{\tilde E^a}(z)}{z}
= \frac{E^a(z)}{z} - \frac{m_a}{M_h}\,D_1^a(z).
\nonumber
\ea
\end{minipage}
}
\newline\newline
The tilde functions that appear in the cross sections are in fact 
precisely the socalled interaction dependent parts of the twist three 
functions. They would vanish in any naive parton model calculation in 
which cross sections are obtained by folding e-parton cross 
sections with parton densities. Considering the relation for $\tilde E$ 
one can state it as $E(z)/z$ = $(m/M_h)\,D_1(z)$ in the absence of 
quark-quark-gluon correlations. The inclusion of the latter also 
require diagrams dressed with gluons (see next section).

In 1-particle inclusive processes, one actually becomes sensitive to 
quark transverse momentum dependent structure functions already for 
unpolarized targets. Comparing with inclusive scattering one has four 
structure functions for unpolarized leptons instead of two 
(longitudinal $W_L$ and transverse $W_T$). One of the additional 
(interference) structure functions ($W_{LT}$) has a azimuthal dependence
$\propto \cos(\phi_h)$. This structure function is order $1/Q$ and 
involves the twist three distribution function $f^\perp$ and the 
fragmentation function $D^\perp$, of the latter again only the 
interaction-dependent part. With polarized leptons (target still 
unpolarized) a fifth structure function can be measured proportional to 
$\sin (\phi_h)$, involving the distribution function $e$ and the 
time-reversal odd fragmentation function $H_1^\perp$.
Explicitly one has~\cite{LM94}
\newline\newline
\fbox{
\begin{minipage}{13.0 cm}  
{\bf Azimuthal asymmetries for unpolarized targets (higher twist)}
\ba
&&\int d^2\bm P_{h\perp}\,\frac{\vert \bm P_{h\perp}\vert}
{M\,z_h} \,cos(\phi_h)
\,\frac{d\sigma_{{OO}}}{d\xbj\,dy\,dz_h\,d^2\bm P_{h\perp}}
%\nonumber \\ &&
= -\frac{2\pi \alpha^2\,s}{Q^4} \,2(2-y)\sqrt{1-y}
\nonumber \\ &&
\hspace{4 cm} \mbox{}\times \sum_{a,\bar a} e_a^2
\Biggl\{
\frac{2M}{Q}\,\xbj^2 {f_1^{\perp(1)a}}(\xbj)\,{D_1^a}(z_h)
\nonumber \\ && \hspace{5.5 cm} \mbox{} 
+\frac{2M_h}{Q}\,\xbj {f_1^a}(\xbj)
\,\frac{{\tilde D^{\perp(1)a}}(z_h)}{z_h}
\Biggr\}
\\ &&
\mbox{note:} \ {\tilde D^{\perp a}}(z) = D^{\perp a}(z) - zD_1^a(z),
\nonumber
\ea
\ba
&&\int d^2\bm P_{h\perp}\,\frac{\vert \bm P_{h\perp}\vert}
{M\,z_h} \,sin(\phi_h)
\,\frac{d\Delta\sigma_{{LO}}}{d\xbj\,dy\,dz_h\,d^2\bm P_{h\perp}}
%\nonumber \\ &&
= \frac{2\pi \alpha^2\,s}{Q^4}\,{\lambda_e}
\,2y\sqrt{1-y}
\nonumber \\ &&
\hspace{4 cm} \mbox{}\times \sum_{a,\bar a} e_a^2
\,\frac{2M}{Q}\,\xbj^2 {\tilde e^a}(\xbj)\,{H_1^{\perp (1)a}}(z_h)
\\ &&
\mbox{note:}
\ {\tilde e^a}(x) = e^a(x) - \frac{m_a}{M}\,\frac{f_1^a(x)}{x}.
\nonumber
\ea
\end{minipage}
}
\newline\newline
Leading order azimuthal asymmetries also exist. They require polarized 
targets and give access to the twist two functions $g_{1T}^{(1)}$ and
$h_{1L}^{\perp (1)}$. These are particularly intersting in view of the 
relations with the twist three functions $g_T$ and $h_L$. One can check 
this relation and/or use these cross sections as an alternate way to find 
these functions. One has~\cite{Collins93,Kotzinian95,TM95b}
\newline \newline
\fbox{
\begin{minipage}{13 cm}  
{\bf Azimuthal asymmetries for polarized targets (leading twist)}
\ba
&&
\int d^2\bm P_{h\perp}\,\frac{\vert \bm P_{h\perp}\vert}
{M\,z_h} \,cos(\phi_h-\phi_S)
\,\frac{d\sigma_{{LT}}}{d\xbj\,dy\,dz_h\,d^2\bm P_{h\perp}}
\nonumber \\ && \hspace{2 cm}
= \frac{2\pi \alpha^2\,s}{Q^4}\,{\lambda_e\,\vert \bSt \vert}
\,y(2-y)\sum_{a,\bar a} e_a^2
\,\xbj\,{g_{1T}^{(1)a}}(\xbj) {D^a_1}(z_h),
\ea
\ba
&&
\int d^2\bm P_{h\perp}\,\frac{\vert \bm P_{h\perp}\vert^2}{MM_h\,z_h^2}
\,sin(2\phi_h)
\,\frac{d\Delta\sigma_{{OL}}}{d\xbj\,dy\,dz_h\,d^2\bm P_{h\perp}}
\nonumber \\ && \hspace{2 cm}
= \frac{2\pi \alpha^2\,s}{Q^4}\,{\lambda}
\,2(1-y)\sum_{a,\bar a} e_a^2
\,\xbj\,{h_{1L}^{\perp(1)a}}(\xbj) {H_1^{\perp(1)a}}(z_h),
\ea
\end{minipage}
}
\newline
\fbox{
\begin{minipage}{13 cm}
\ba
&&
\int d^2\bm P_{h\perp}\,d\phi_e
\,\frac{\vert \bm P_{h\perp}\vert}{M_h\,z_h}
\,sin(\phi_h+\phi_S)
\,\frac{d\Delta\sigma_{{OT}}}
{d\xbj\,dy\,dz_h\,d\phi_e\,d^2\bm P_{h\perp}}
\nonumber \\ && \hspace{2 cm}
= \frac{2\pi \alpha^2\,s}{Q^4}\,{\vert \bSt \vert}
\,2(1-y)\sum_{a,\bar a} e_a^2
\,\xbj\,{h_1^a}(\xbj) {H_1^{\perp(1)a}}(z_h).
\ea
\end{minipage}
}
\newline \newline
Note that these are leading contributions in the cross section. At 
subleading order the same asymmetries may have higher harmonics, such as
a $\sin(3\phi_h -\phi_S)$ asymmetry appearing in $\Delta \sigma_{OT}$
$\sim$ ${h_{1T}^{\perp (2)}}{H_1^{\perp(1)}}$.

I want to end this section with giving an indication of the magnitude of 
a $\bkt$-dependent distribution functions, in particular look at 
$g_{1T}^{(1)}(x)$. For this we use the relation of this function with 
$g_2$, namely $g_{1T}^{(1)}(x)$ = $-\int_x^1 dy\,g_2(y)$ and the SLAC 
E143 data on $g_2$. The result is shown in Fig. 4 (left panel). Except 
for the data also the result obtained from the Wandzura-Wilczek part of 
$g_2$, namely $g_2^{WW}(x)$ = $-g_1(x) + \int_x^1 dy\,(g_1(y)/y)$ 
(omitting quark mass terms). This result is shown in Fig. 4 (left panel) 
as the solid line~\cite{KM96}. 
\begin{figure}[hb]
\leavevmode
\epsfxsize=6.5 cm \epsfbox{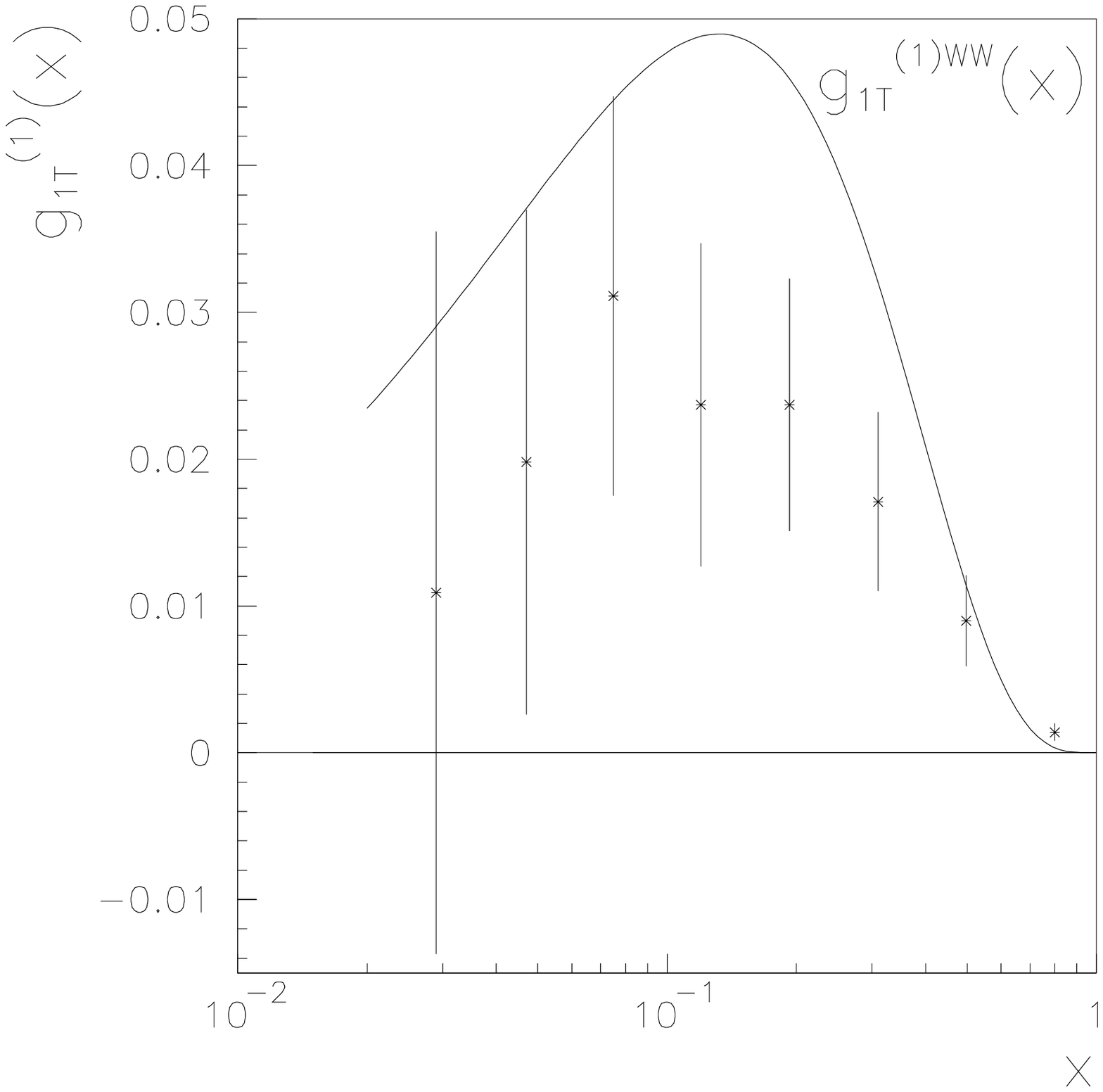}
\hspace{0.5 cm}
\epsfxsize=6.5 cm \epsfbox{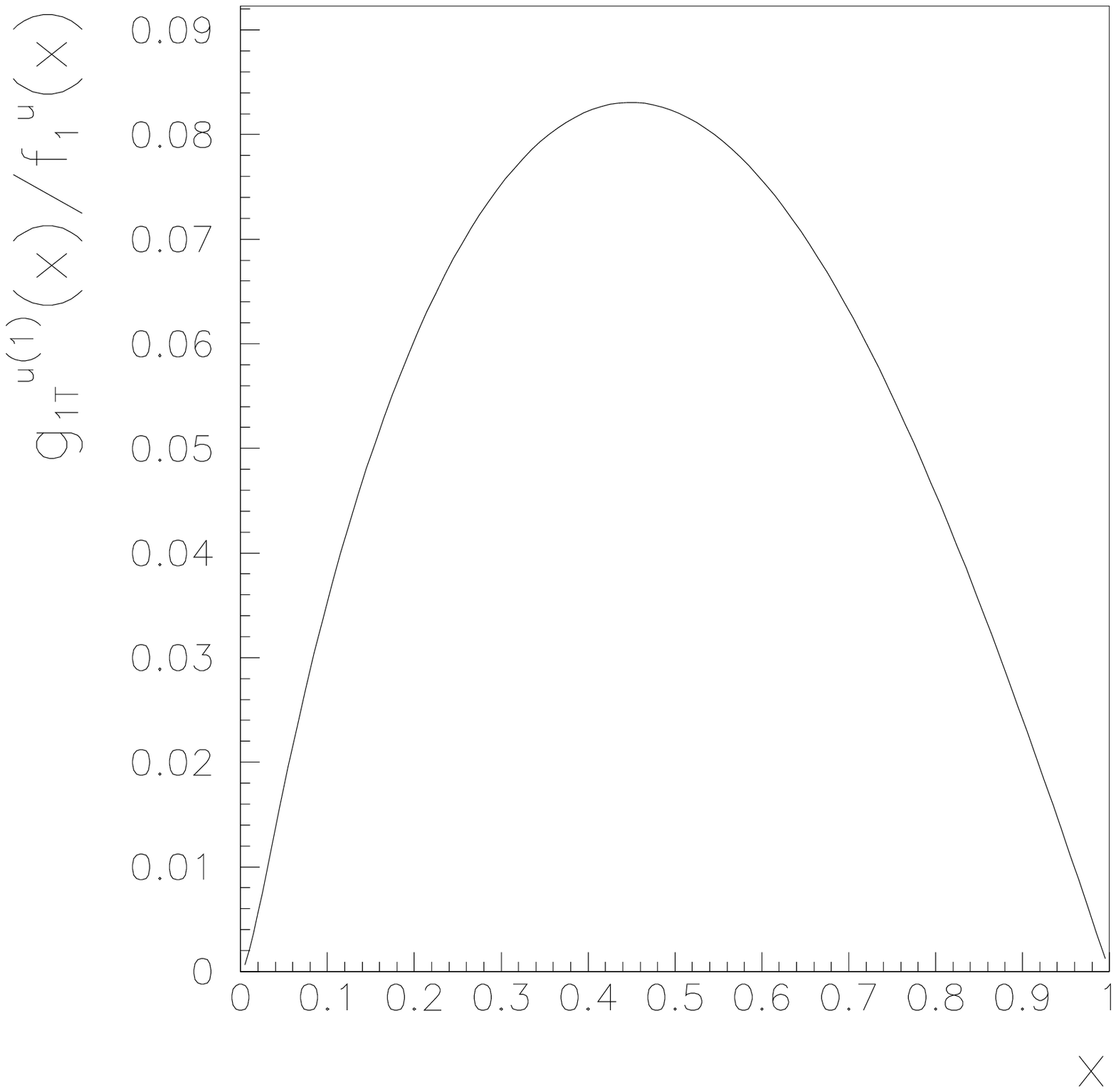}
\caption{\label{fig4}\em
Left panel: the function $g_{1T}^{(1)}$ estimated from the E143 data on 
$g_2$ (datapoints) and from the Wandzura-Wilczek part of $g_2$ (solid line);
right panel: the resulting asymmetry in 
$ep \rightarrow e\pi^+ X$.}
\end{figure}
\newline
The resulting asymmetry in $\sigma_{LT}$ is except for kinematical 
factors depending on $y$ proportional to the ratio 
$g_{1T}^{(1)}(x)/f_1(x)$. For the case of $\pi^+$ production in polarized 
lepton scattering off a transversely polarized proton, it is dominantly  
the u-quark that contributes. This ratio is shown in the right panel of 
Fig. 4.

\section{The full calculation and concluding remarks}

In the previous section the lowest order results of the calculations of 
1-particle inclusive lepton-hadron scattering have been presented. What 
other effects are important in these cross sections. As emphasized 
before, QCD is believed to provide a reliable framework to obtain the 
soft parts. Within this framework we can also indicate some
theoretical and experimental caveats.

For this, lets write down the first few diagrams in the full calculation 
within the field theoretical approach
\newline\newline
\begin{minipage}{6.0 cm}
\leavevmode
\epsfxsize=5 cm \epsfbox{mulders2.eps}
\end{minipage}
\begin{minipage}{7.5 cm}
The first contribution is the parton model result, giving the expressions 
as derived in the previous section. It is important to realize that its 
validity requires $P\cdot P_h \gg M^2$, the {\em current fragmentation 
region} in which there is a sufficiently large 
rapidity gap between target remnant and produced particles.
\end{minipage}
\newline
\begin{minipage}{7.5 cm}
The case that $P\cdot P_h \sim M^2$, the {\em target fragmentation region} 
involves by the definition of soft parts, those soft parts that contain 
both target and produced hadron (fracture functions)~\cite{TV93}. 
These parts need a separate theoretical treatment. The use of the results 
derived previously is limited to the current fragmentation region.
\end{minipage}
\begin{minipage}{6.0 cm}
\leavevmode
\epsfxsize=5 cm \epsfbox{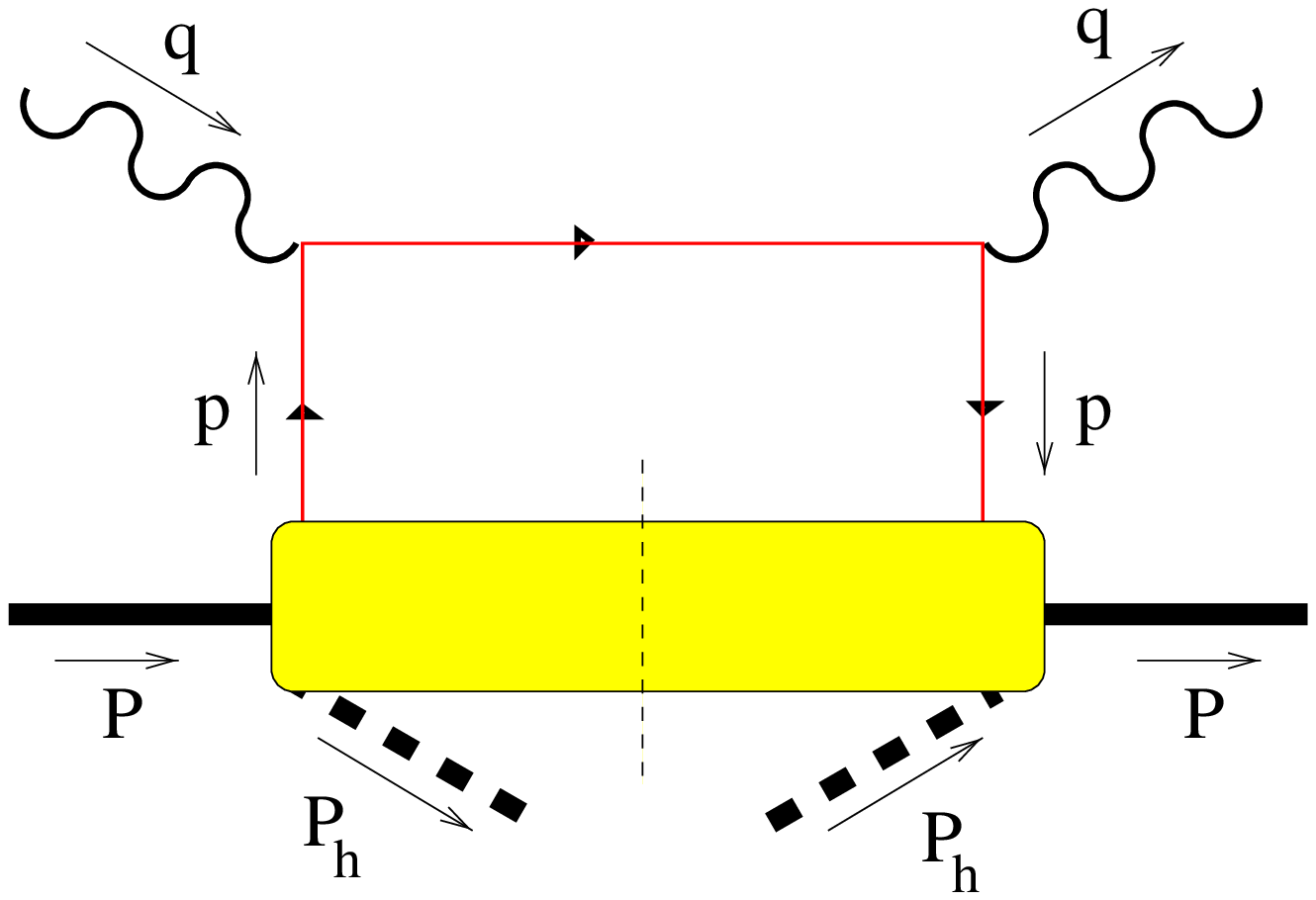}
\end{minipage}
\newline
Diagrams containing gluons attached to the soft part (one example shown)
need to be considered carefully. They are of two types:
\newline
\begin{minipage}{6.0 cm}
\leavevmode
\epsfxsize=5 cm \epsfbox{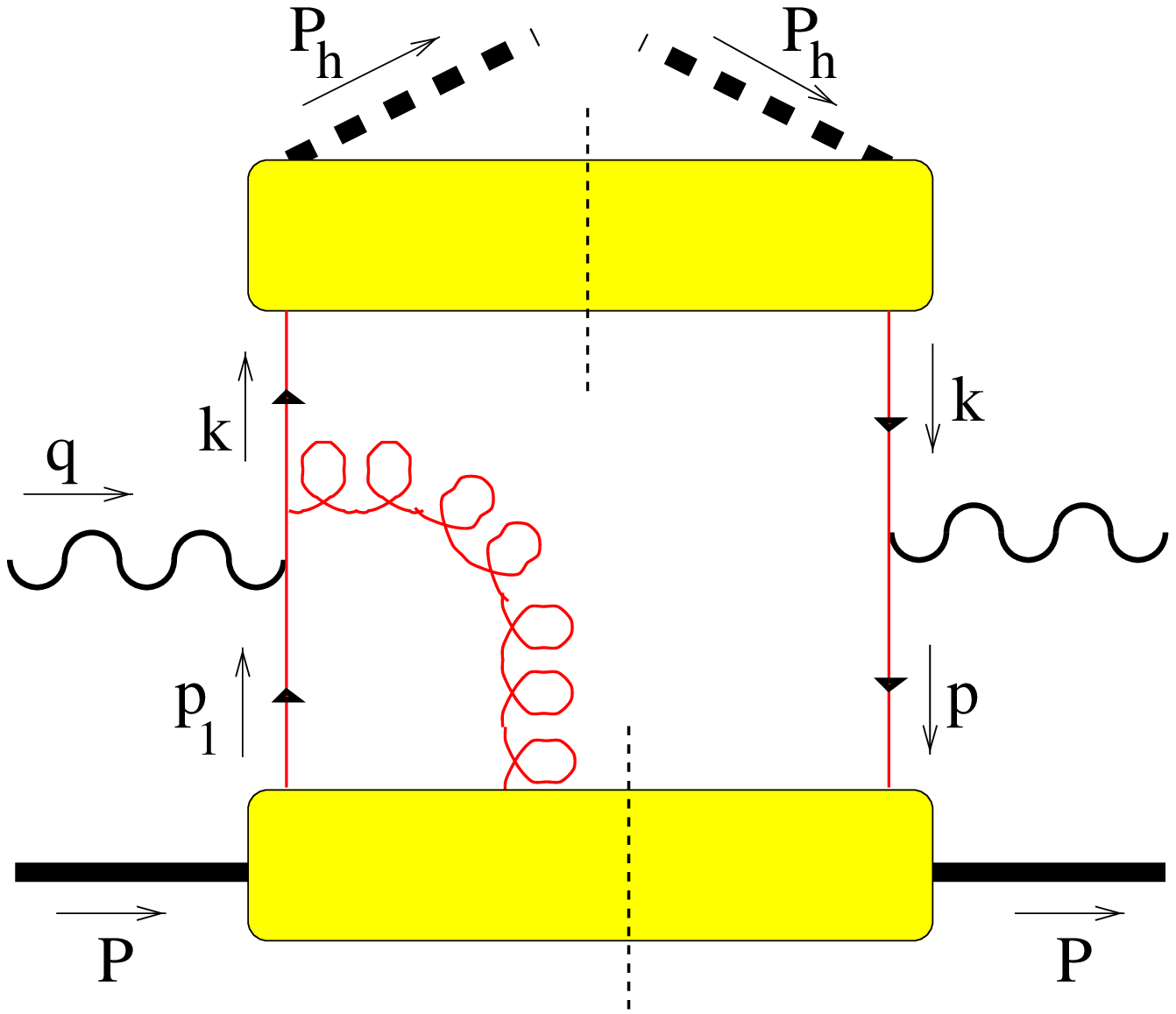}
\end{minipage}
\begin{minipage}{7.5 cm}
\begin{itemize}
\item
Longitudinal $A^+$ ($A^-$) gluons. They do contribute at leading order, but 
can be absorbed into the definition of the soft parts, creating the link 
operator that renders the definition of $\Phi$ ($\Delta$) color gauge invariant.
\item
Transverse $A_T^\alpha$ gluons. They contribute at order $1/Q$. They do 
not lead to new independent distribution (fragmentation) functions, courtesy
of the QCD equations of motion. They ensure electromagnetic gauge 
invariance of the full calculation at order $1/Q$.
\end{itemize}
\end{minipage}
\newline
The next corrections that we want to consider are quark and gluon ladders.
They give the asymptotic $\bpt$-dependence of the distribution
functions, which turns out to be proportional to $\alpha_s/\bpt^2$. 
\begin{center}
\leavevmode
\epsfxsize=5 cm \epsfbox{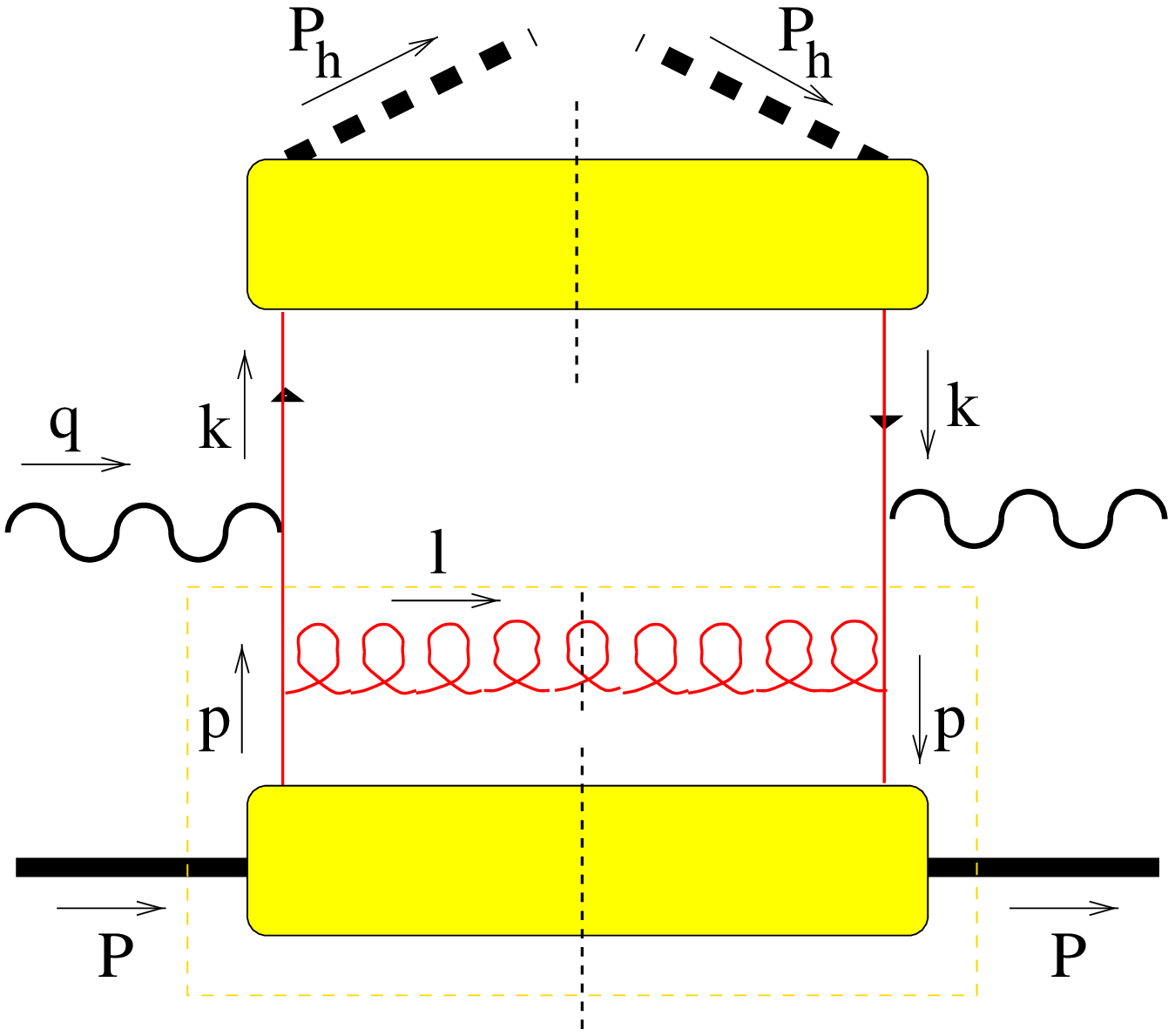}
\hspace{1 cm}
\epsfxsize=5 cm \epsfbox{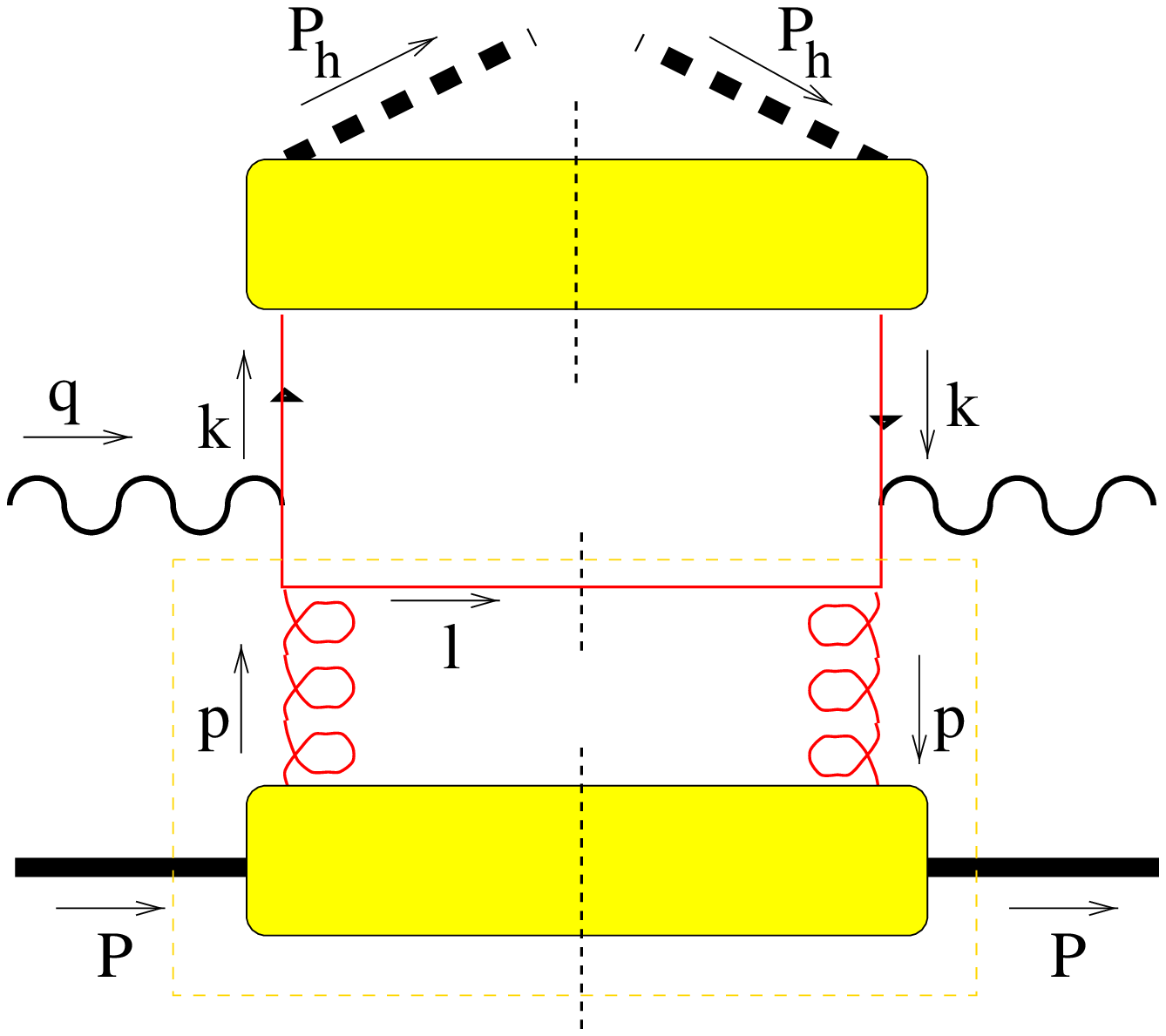}
\end{center}
For the integrated 
distribution functions, this leads to $\alpha_s \log(Q^2/\mu^2)$ contributions.
These can be incorporated into a scale dependence of the distribution 
functions, described with the GLAP equations.
\newline
\begin{minipage}{6.0 cm}
\leavevmode
\epsfxsize=5 cm \epsfbox{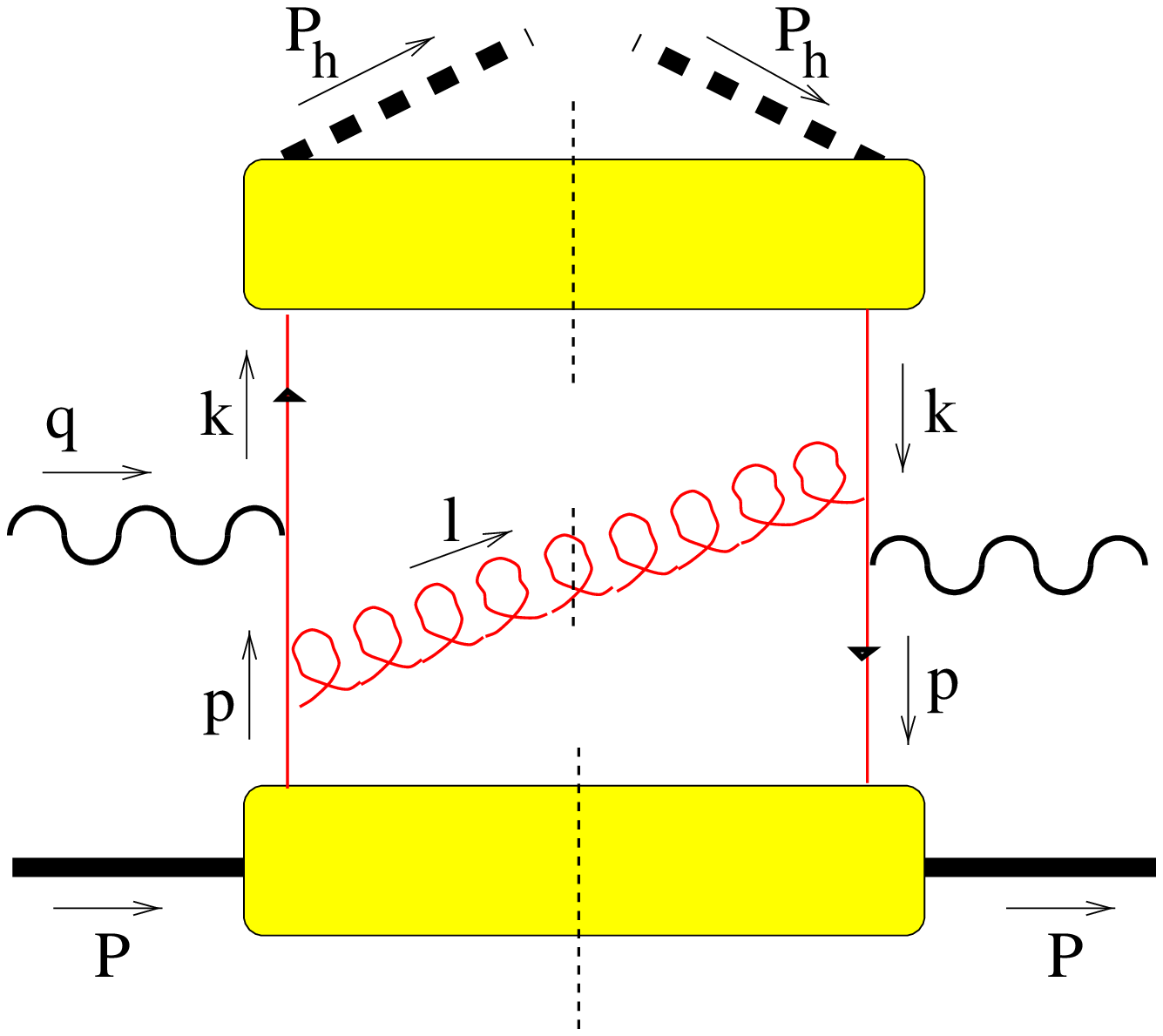}
\end{minipage}
\begin{minipage}{7.5 cm}
The final QCD corrections which we discuss are process dependent $\alpha_s$
corrections, for 1-particle inclusive scattering arising from a diagram
as shown here. A characteristic example for inclusive DIS is the 
longitudinal part,
\begin{eqnarray*}
F_L = \frac{F_2}{x}-2F_1 \sim \alpha_s(Q^2)\times\ldots.
\end{eqnarray*}
\end{minipage}

In my talk I have tried to present results that show the prospects 
of spin physics in semi-inclusive, in particular 1-particle inclusive 
lepton-hadron scattering. The goal is the study of the quark
and gluon structure of hadrons, emphasizing the spin structure and the
dependence on quark transverse momenta. The reason why the prospect is
promising is the existence of a field theoretical framework that allows
a clean study involving well-defined hadronic matrix elements. 

This work is part of the scientific program of the 
foundation for Fundamental Research on Matter (FOM),
the Dutch Organization for Scientific Research (NWO) 
and the TMR program ERB FMRX-CT96-0008.

\end{document}